\documentstyle[psfig,prl,aps,multicol]{revtex}

\begin{document}
\draft                                           

\title{   Dynamical Mean-Field Theory for Doped Antiferromagnets }

\author{  Marcus Fleck } 
\address{ Max-Planck-Institut f\"{u}r Festk\"{o}rperforschung, 
          Heisenbergstrasse 1, D-70569 Stuttgart, Germany } 

\author{  Alexander I. Lichtenstein } 
\address{ Forschungszentrum J\"{u}lich, D-52425 J\"{u}lich, Germany }

\author{  Andrzej M. Ole\'s\cite{AMO} and Lars Hedin }
\address{ Max-Planck-Institut f\"{u}r Festk\"{o}rperforschung, 
          Heisenbergstrasse 1, D-70569 Stuttgart, Germany } 

\author{  Vladimir I. Anisimov   }
\address{ Institute of Metal Physics, Russian Academy of Sciences,
          620219 Ekaterinburg, GSP-170, Russia }

\date{30 May 1997}
\maketitle

\begin{abstract}
We have generalized the dynamical mean-field theory to study the doping 
dependence of the crossover from antiferromagnetic to short-range order 
modelled by an incommensurate spin density wave in the Hubbard model. 
The local selfenergy which includes spin fluctuations gives quasiparticle 
weights and spectral properties in good agreement with quantum Monte Carlo 
and exact diagonalization data in two dimensions. 
The spectra at finite doping are characterized by a Mott-Hubbard `gap'
accompanied by a pseudogap induced by the local spin order.  
\end{abstract}


\begin{multicols}{2} 

Although an important progress in the present understanding of strongly 
correlated fermion systems occurred recently due to numerical methods, such
as quantum Monte Carlo (QMC) and exact diagonalization (ED), an analytic 
treatment that maintains local correlations is needed to investigate the 
thermodynamic limit. An attractive possibility is the limit of large spatial 
dimension ($d=\infty$), when the diagrams in the perturbative expansion 
collapse to a single site and the fermion dynamics is described by a {\em 
local selfenergy\/} \cite{Met89}. This allows to map lattice models onto 
quantum impurity models, which can then be solved self-consistently in the 
dynamical mean-field theory (DMFT) \cite{Geo96}. At large Coulomb interaction 
$U$, the spectral properties of Mott insulators were reproduced using the 
DMFT both at half- \cite{Geo96}, and at arbitrary filling $n$ \cite{Kaj96}, 
but magnetic long-range order (LRO) 
was analyzed at $d=\infty$ only at half-filling ($n=1$) \cite{Log96}. 
  
The modification of the magnetic order under doping $\delta=1-n$ away from 
$n=1$ is a central issue in the theory of the oxide superconductors. Undoped 
La$_2$CuO$_4$ is a {\it commensurate} antiferromagnetic (AF) insulator, 
while doping by Sr into La$_{2-x}$Sr$_x$CuO$_4$ results in short-range AF 
order within {\it incommensurate} magnetic structures \cite{Tra97}. The 
physical properties of doped CuO$_2$ planes of the high-$T_c$ superconductors 
can be successfully described near $n=1$ using a two-dimensional (2D) 
single-band Hubbard (or $t-J$ \cite{Dag94}) model, and indeed incommensurate 
magnetic order was found analytically \cite{Shr89}, in Hartree-Fock (HF) 
\cite{Zaa89}, and in slave-boson approximation \cite{Kan90,Arr91}.
Motivated by the successful description of the transport properties of the
CuO$_2$ planes within the DMFT \cite{Pru95,Geo96}, we present in this Letter 
a generalization of the DMFT to the magnetically ordered states, using 
the Berk-Schrieffer \cite{Ber66} spin-fluctuation exchange interaction with 
an effective potential due to particle-particle scattering \cite{Che91}, and 
construct a selfenergy which leads to an overall good agreement with the QMC 
and ED data. It demonstrates a combination of physics arising from the Slater 
picture and the Mott-Hubbard description of strongly correlated electron 
systems.  

In order to simulate the weakened short-range AF order in the doped systems, 
we consider an incommensurate spin spiral (SS) structure. The direction of 
the magnetization $m$ at site $i$, with respect to the global $z$-axis is 
specified by two spherical angles, $\Omega_i=(\phi_i,\theta_i)$. 
Thus, we transform the original fermions
$\{a_{i\uparrow}^{\dagger},a_{i\downarrow}^{\dagger}\}$ to the fermions 
quantized with respect to local quantization axis at each site \cite{Arr91},
$c^{\dagger}_{i\sigma}=\sum_{\lambda}\left[
   {\cal R}(\Omega_i)\right]_{\sigma\lambda}a^{\dagger}_{i\lambda}$, where
${\cal R}(\Omega_i)=e^{-i(  \phi_i/2)\hat{\sigma}_z}
                    e^{-i(\theta_i/2)\hat{\sigma}_y}$, 
and $\hat{\sigma}_y$ and $\hat{\sigma}_z$ are Pauli spin matrices. 
This transforms the Hubbard Hamiltonian with 
nearest-neighbor hopping, $t_{ij}=-t$, to
\begin{equation}
H\!=\!\sum_{ij,\sigma\sigma'}\!\!t_{ij}c^{\dagger}_{i\sigma}
\left[{\cal R}^*(\Omega_i){\cal R}(\Omega_j)\right]_{\sigma\sigma'}
c^{}_{j\sigma'}\! + U\! \sum_i n_{i\uparrow}n_{i\downarrow} .
\label{hubbard}
\end{equation}

We take the polar angle as site-independent, $\theta_i=\theta$, and let 
the azimuthal angle describe a spin spiral with a wave-vector ${\bf Q}$, 
$\phi_i={\bf Q}\cdot{\bf R}_i$. By a straightforward diagonalization of 
the kinetic energy one finds the energies,
\begin{eqnarray}
\hat{T}_{\bf Q}({\bf k})
&=&\frac{1}{2}\;\varepsilon_{{\bf k}-\frac{\bf Q}{2}}
  (\hat{1}+\cos\theta\;\hat{\sigma}_z-\sin\theta\;\hat{\sigma}_x) \nonumber \\
&+&\frac{1}{2}\;\varepsilon_{{\bf k}+\frac{\bf Q}{2}}
  (\hat{1}-\cos\theta\;\hat{\sigma}_z+\sin\theta\;\hat{\sigma}_x),
\label{ekin}
\end{eqnarray}
where $\varepsilon_{\bf k}=-2t(\cos k_x+\cos k_y)$. 
We limit ourselves to plane spirals \cite{notespirals}, and choose 
$\theta=\pi/2$. 

The one-particle Green function is described by a $(2\times 2)$ matrix 
$\hat{G}_{\bf Q}({\bf k},i\omega_{\nu})$ in spin space, where 
$\omega_{\nu}$ are fermionic Matsubara frequencies. We approximate the 
Green function using a {\it local selfenergy} \cite{Met89,Geo96},
\begin{equation}
\hat{G}^{-1}_{\bf Q}({\bf k},i\omega_{\nu})=i\omega_{\nu}+\mu 
- \hat{T}_{\bf Q}({\bf k}) - \hat{\Sigma}^{\rm HF}_{\bf Q}
- \hat{\Sigma}_{\bf Q}^{\rm SF}(i\omega_{\nu}).
\label{localg}
\end{equation}
The local lattice Green function, 
$\hat{G}_{\bf Q}(i\omega_{\nu})=N_k^{-1}\sum_{\bf k}
\hat{G}_{\bf Q}({\bf k},i\omega_{\nu})$, becomes diagonal due to the parity 
of $\hat{T}_{\bf Q}({\bf k})$. Therefore, all local 
quantities including the selfenergy $\Sigma$, which consists of the HF part,
${\Sigma}^{\rm HF}_{{\bf Q}\sigma}=U\langle n_{0\bar{\sigma}}\rangle$ 
with $\bar{\sigma}=-\sigma$, and the spin-fluctuation (SF) part, 
$\Sigma_{{\bf Q}\sigma}^{\rm SF}(i\omega_{\nu})$, are diagonal. Using the 
{\it cavity method\/} \cite{Geo96} for a hypercubic lattice at $d=\infty$, 
we showed that the dynamical Weiss field, 
${\cal G}^{0}_{{\bf Q},\sigma}(i\omega_{\nu})$, can be computed from the 
Dyson equation of the spin-symmetry broken Anderson impurity model, 
\begin{equation}
\hat{\cal G}^{0}_{\bf Q}(i\omega_{\nu})\,^{-1}=
\hat{G}_{\bf Q}(i\omega_{\nu})\,^{-1}
+\hat{\Sigma}_{\bf Q}^{\rm SF}(i\omega_{\nu})\;.
\label{cavity}
\end{equation}

The last and the most important step is to find an appropriate expression
for selfenergy. This is known to be notoriously difficult and the usual 
approach within the iterative perturbation scheme (IPS) \cite{Kaj96} is to 
develop an ansatz which reproduces the exact results in certain limits. In 
spite of the correct large $U$ limit \cite{Kaj96}, the absence of a reliable 
form of $\Sigma$ hindered further applications of the DMFT for the magnetic 
states in the intermediate $U$ range. By considering the Kadanoff-Baym 
potential containing a class of diagrams up to infinite order \cite{Fle97}, 
we constructed the SF part of selfenergy, 
\begin{eqnarray}
\Sigma_{{\bf Q}\sigma}^{\rm SF}(i\omega_{\nu})        
&=&{\bar U}^{2}\beta^{-1} \sum_{\mu} 
\chi_{\bar{\sigma}\sigma{\bf Q}}(i\omega_{\mu})\, 
{\cal G}^{0}_{{\bf Q}\bar{\sigma}}(i\omega_{\nu}-i\omega_{\mu}) \nonumber \\ 
 & + & {\bar U}^{2}\beta^{-1} \sum_{\mu} 
\chi_{\bar{\sigma}\bar{\sigma},{\bf Q}}(i\omega_{\mu})\, 
{\cal G}^{0}_{{\bf Q}\sigma}(i\omega_{\nu}-i\omega_{\mu}),           
\label{sigma}
\end{eqnarray}
with $\beta=1/k_BT$. Here we approximated $\Sigma[G]$ by $\Sigma[{\cal G}^0]$.
Eq. (\ref{sigma}) resembles the selfenergy derived by the coupling of the 
moving hole to transverse spin-fluctuations \cite{Alt95}, but we note that 
the coupling to longitudinal spin-fluctuations gives a significant 
contribution. The selfenergy in a magnetic system is calculated using 
the symmetry-broken magnetic HF state. The transverse, 
$\chi_{\bar{\sigma}\sigma}={\bar U}[\chi^{(0)}_{\bar{\sigma}\sigma}]^2
 /[1-{\bar U}\chi^{(0)}_{\bar{\sigma}\sigma}]$, and longitudinal, 
$\chi_{\sigma\sigma}=\chi^{(0)}_{\sigma\sigma}/[1-{\bar U}^2
 \chi^{(0)}_{\uparrow\uparrow}\chi^{(0)}_{\downarrow\downarrow}]$,
susceptibility in Eq. (\ref{sigma}) are found in random phase approximation 
(RPA) with renormalized interaction $\bar{U}$. Thus, the renormalized 
interaction $\bar U$ is not a fitting parameter \cite{Bul93}, but results 
from the screening by particle-particle diagrams \cite{Che91,Fle97}, 
\begin{equation}
\label{ubar}
\bar{U}=U/[1+U\chi^{pp}(0)], 
\end{equation}
where 
$\chi^{pp}(0)=\beta^{-1}
\sum_{\mu}{\cal G}^{0}_{{\bf Q}  \uparrow}( i\omega_{\mu})\, 
          {\cal G}^{0}_{{\bf Q}\downarrow}(-i\omega_{\mu})$. 
This screening effect gives the magnetic structure factor \cite{Che91} and 
the selfenergy \cite{Bul93} calculated from Eq. (\ref{sigma}) in good 
agreement with the QMC results, and depends on the underlying magnetic order. 
It is largest in the paramagnetic states and vanishes in the N\'eel state at 
$n=1$ for $U\to\infty$, and is thus very important when the magnetic phase 
diagrams are considered \cite{Fle97}. 

Eqs. (\ref{localg}), (\ref{cavity}), (\ref{sigma}), and (\ref{ubar}) 
represent a solution for the one-particle Green function within the DMFT in 
the IPS. They have been solved self-consistently and the energetically 
stable spiral configuration was found. This procedure is further justified
by the sum rule \cite{Vil97},
\begin{equation}
\label{sumrule}
\frac{1}{2\beta}\sum_{\nu\sigma}
     \Sigma_{\sigma}(i\omega_{\nu})G_{\sigma}(i\omega_{\nu})
     e^{i\omega_{\nu}0^+}
     =U\langle n_{0\uparrow}n_{0\downarrow}\rangle, 
\end{equation}
which is well fulfilled in the present approach with 
$U\langle n_{0\uparrow}n_{0\downarrow}\rangle\simeq \bar{U}
  \langle n_{0\uparrow}\rangle\langle n_{0\downarrow}\rangle$ \cite{Fle97}. 
We also show below that the local selfenergy (\ref{sigma}) leads to an 
overall satisfactory agreement with the QMC and ED data. 
\begin{figure}
\centerline{\psfig{figure=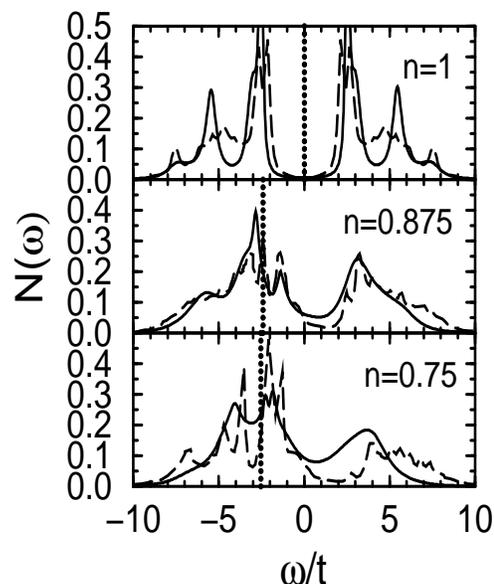,height=3.0in,width=2.5in}}
\narrowtext
\smallskip
\caption
{Total densities of states $N(\omega)$ as obtained within DMFT (solid
 lines) for $\delta=0$ (AF state), 0.125 [(1,1) spiral], and 0.25 [(1,0)
 spiral] with $U/t=8$ and $T=0.05t$. The dashed lines reproduce $N(\omega)$
 found by ED of a $4\times 4$ lattice in Ref. \protect{\cite{Dag92}}, and the 
 dotted lines show the Fermi energy. }
\label{dos}
\end{figure}

The system with local selfenergy is not fully 2D, and calculations performed 
at low temperature $T=0.05t$ ($T\ll J=4t^2/U$) converge to states with 
magnetic LRO, and we discuss here the case of $U/t=8$. Away from half-filling 
one finds incommensurate states characterized by 
${\bf Q}=[\pi(1-2\eta_x),\pi(1-2\eta_y)]$, $\eta_x=\eta_y\neq 0$ for 
SS(1,1) and $\eta_x\neq 0$, $\eta_y=0$ for SS(1,0), 
the neighboring spins are not strictly antiparallel, the kinetic energy 
$\sim t\sin 2\pi\eta_{\alpha}$ is gained in agreement with polaron physics, 
and the dynamics becomes closer to free electrons. We reproduce a typical 
pattern found in HF and in slave-boson calculations \cite{Kan90,Arr91}: 
First, the AF state changes to a mixed AF-SS(1,1) state in the range of  
$0<\delta<0.116$, which indicates that a domain-wall phase \cite{Zaa89} 
might be more stable at low doping. Next, a first order transition to a 
SS(1,0) is found at $\delta\simeq 0.23$, and both spirals coexist in the 
range $0.203<\delta<0.248$. The moments decrease under doping, and finally 
disappear continuosly at $\delta\simeq 0.65$. 

Unlike in the IPS based on second order perturbation theory \cite{Geo96}, 
the AF order [${\bf Q}=(\pi,\pi)$] survives at $n=1$ for $U>4t$, as found
in QMC at $d=\infty$ \cite{Ulm95}. 
In agreement with experiment \cite{Wel95}, the spectral functions 
at $n=1$, $A({\bf k},\omega)=-\pi^{-1}\sum_{\sigma\sigma'}{\rm Im} 
G_{{\bf Q},\sigma\sigma'}({\bf k}-{\bf Q}/2,\omega+i\epsilon)$, have the 
same generic shape as in $t-J$ model \cite{Dag94,Ste91}. The photoemission 
spectrum consists of a {\it coherent\/} quasiparticle (QP) state near the 
gap with a dispersion $\sim 2J$, and an incoherent part of width $\sim 7t$ 
at lower energies. The DMFT reproduces the nesting instability towards the 
AF order at $n=1$, but the AF gap is considerably reduced. At $U/t=8$ the HF 
gap of $7.14\ t$ is reduced to $\Delta\simeq 4.93\ t$, and agrees very well 
with the ED data for the $4\times 4$ cluster \cite{Dag92} (see Fig. 
\ref{dos}). In contrast, the gap reduction is overestimated within 
approaches \cite{Alt95} not treating properly local correlations 
(\ref{sumrule}). 
\begin{figure}
\centerline{\psfig{figure=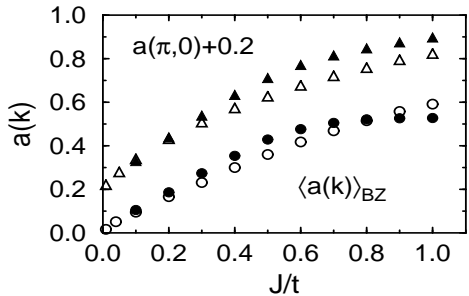,height=1.5in,width=2.4in}}
\narrowtext
\smallskip
\caption
{Spectral weight of a coherent QP at ${\bf k}=(\pi,0)$, 
 $a(\pi,0)+0.2$ (triangles), and the averaged weight over Brillouin zone,
 $\langle a({\bf k})\rangle_{\rm BZ}$ (circles), 
 as obtained at $n=1$ in DMFT (full symbols) and in SCBA (empty symbols) 
 of Ref. \protect{\cite{Ger91}}.}
\label{weight}
\end{figure}

The QP weight $a({\bf k})$ increases with increasing $J/t$, in agreement 
with the self-consistent Born approximation (SCBA) (and ED data) \cite{Ger91} 
in the range $J/t<0.5$, both at the $X=(\pi,0)$ point, and for the averaged 
weight $\langle a({\bf k})\rangle_{\rm BZ}$ (Fig. \ref{weight}). The latter 
first increases with $J/t$, as in the $t-J$ model, but then flattens out 
around $J/t\simeq 0.8$, and starts to saturate at a value lower than one. 
This behavior follows in first place from the absence of the intersite SFs, 
but also shows that the $t-J$ model does not apply for $J/t>0.8$, where the 
charge excitations to the upper Hubbard band (UHB) become important. 

The quality of our approach is demonstrated by the overall shapes of the 
densities of states $N(\omega)=N_k^{-1}\sum_{\bf k}A({\bf k},\omega)$ which 
are in a very good agreement with the ED data not only at half-filling, but 
also for doped systems (Fig. \ref{dos}). At finite doping the
system becomes metallic, and the selfenergy (\ref{sigma}) exhibits a 
Fermi-liquid behavior, with the imaginary part $\sim (\omega-\mu)^2$ 
vanishing at the Fermi energy $\mu$. This result might change, however, in
a real system due to critical scattering mediated by spin-waves. The spectral 
weight is transferred from the UHB into the inverse photoemission (IPES) part 
of the lower Hubbard band (LHB), and $N(\omega)$ agrees very well with the ED 
for the SS(1,1) ground state with $\eta_x=\eta_y\simeq \delta$ at 
$\delta=0.125$, both for the positions of the UHB and LHB, and their 
intensities. We reproduce thereby the increase of the IPES spectral weight 
above $2\delta$, being $0.331$ at $U/t=8$ \cite{Esk94}. A somewhat smaller 
distance between the LHB and the UHB for the SS(1,0) at $\delta=0.25$ might 
suggest that the system properties are closer to a disordered local moment 
state.
\begin{figure}
\centerline{\psfig{figure=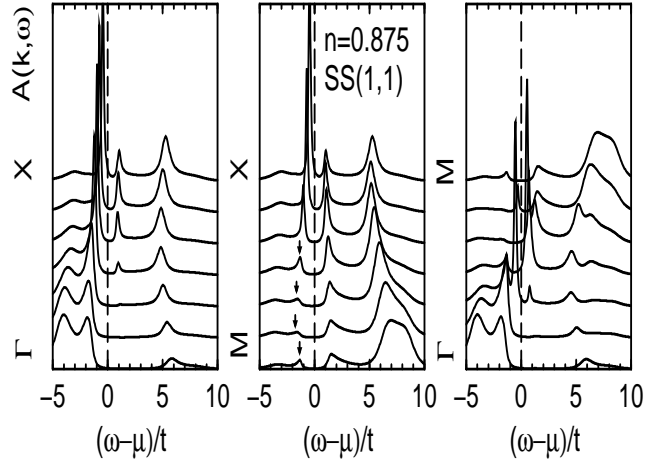,height=2.4in,width=3.3in}}
\narrowtext
\smallskip
\caption
{Spectral functions $A({\bf k},\omega)$ in the main BZ directions: $\Gamma-X$,
 $M-X$, and $\Gamma-M$ in SS(1,1) state at $\delta=0.125$ and $T=0.05t$. The 
 spectra along the $\Gamma-M$ direction have been averaged over the (1,1) 
 and ($\bar{1}$,$\bar{1}$) spirals. A shadow band in $M-X$ direction is 
 indicated by arrows.}
\label{ak}
\end{figure}

The ${\bf k}$-resolved spectral functions $A({\bf k},\omega)$ allow to 
clarify the generic features of the doped systems: a large Mott-Hubbard 
gap persists between the LHB and UHB (with energies $\sim -3.3t$, $\sim 6.2t$
in Fig. \ref{ak}, respectively), while a new smaller {\it pseudogap\/} forms 
between the photoemission and IPES parts of the LHB \cite{notepseudo}. 
The latter is induced 
by the underlying spiral magnetic order which removes the degeneracy between 
$X$ and $(\pi/2,\pi/2)$ points at $\delta=0.125$, and separates two features 
which originate from a single QP peak at $n=1$; the one with a higher 
intensity indicates the position of the QP, while the other represents a 
shadow band. These structures and the pseudogap are best visible in the 
$X-M$ direction. At the $X$ point one finds the QP peak at energy $\simeq 
-0.44t$ and a pseudogap which persists along the $X-M$ direction, while the 
spectral weight moves gradually from the lower to the upper maximum 
\cite{Ede97}. Thus, the QP peak at $X$ point tranforms gradually into a 
shadow band of width $\sim 2J$ when the $M$ point is approached (see Fig. 
\ref{ak}). This interpretation is supported by an excellent agreement with 
the ED data of Ref. \onlinecite{Dag92}. The QP intensities found at 
$\delta=0.125$ are somewhat reduced from those in the AF state at $n=1$, 
and agree qualitatively with ED for the $t-J$ model at the same doping 
\cite{Ste91}. Although we cannot attempt a quantitative comparison within 
the $t-U$ model, our
results agree qualitatively with the QP band with dispersion $\sim 2J$ 
and pseudogap observed in Bi$_2$Sr$_2$CaCu$_2$O$_{8+\delta}$ \cite{Wel95}.
\begin{figure}
\centerline{\psfig{figure=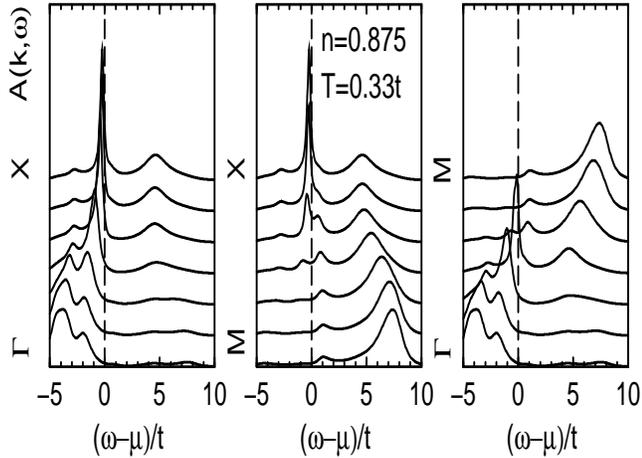,height=2.4in,width=3.3in}}
\narrowtext
\smallskip
\caption
{Spectral functions $A({\bf k},\omega)$ at $\delta=0.125$ and $T=0.33t$.
 The conventions are the same as in Fig. 3.}
\label{akt}
\end{figure}

The dependence of the QPs on the spiral order explains why the ED results
obtained for relatively small clusters \cite{Dag94} cannot reproduce the 
experimental observations that the QP peak at $X$ point moves above  
$\omega=\mu$ with increasing doping \cite{Mar96}. Furthermore, the pseudogap 
gradually disappears with increasing temperature, and the spectral weight 
crosses the Fermi energy along the $X-M$ direction, as found at $T=0.33t$ 
(Fig. \ref{akt}). This is consistent with the recent QMC calculations 
\cite{Pre97}, suggesting that the pseudogap $\sim J$ originates from 
short-range 
magnetic correlations at $T\to 0$. The spectral functions at $T=0.5t$ have 
only broad maxima which correspond to the LHB and UHB, and agree remarkably 
well with the QMC results of Bulut {\it et al.} \cite{Bul94}. At $\delta=0.25$ 
the spectra consist of incoherent processes even at low $T$, accompanied by 
new dispersive features $\sim t$, which demonstrates that the pseudogap and 
the hole-spin correlations are gradually lost in the overdoped regime.

Summarizing, we have developed a simple but very powerful analytic method 
which captures the essential changes of the electronic structure, and gives 
coherent QPs by dressing a moving hole with SFs. While the large Mott-Hubbard 
gap remains almost unchanged with increasing doping, the QP states split into 
two features separated by a pseudogap. 
We believe that the electronic properties are generic and not changed by 
domain-wall formation \cite{Zaa89}. It is expected that the present 
extension of DMFT would significantly improve the understanding of the 
strongly correlated transition metal oxides, if implemented into realistic 
band structure calculations.

We thank P. Horsch, G. Kotliar, A. O. Anokhin, and A. I. Poteryaev for 
valuable discussions, and acknowledge the support by the Committee of 
Scientific Research (KBN) of Poland, Project No.~2 P03B 118 13 (AMO), and by 
the Russian Foundation for Basic Research (RFFI), Grant No.~96-02-16167 (VIA).

\end{multicols} 


\begin{thebibliography}{99}
\bibitem{notespirals} 
                Double spirals were shown to be unstable in the 2D $t-J$
                model in Ref. \protect{\cite{Kan90}}.
\end{thebibliography}
\end{document}